\documentclass[10pt]{article}
\usepackage{graphicx}
\usepackage{dcolumn}
\usepackage{bm}

\begin{document}

\title{Experimental tests on the lifetime asymmetry}
\author {Zhi-Qiang Shi\thanks{E-mail address: zqshi@snnu.edu.cn}\\{\it \small Department of Physics, Shaanxi Normal University, Xi'an 710062, China}\\[0.3cm]
         Guang-Jiong Ni\thanks{E-mail address: pdx01018@pdx.edu}\\
            {\it \small Department of Physics, Fudan University, Shanghai 200433, China}\\
            {\it \small Department of Physics, Portland State University, Portland OR 97207, USA}}
\date{}
\maketitle

\begin{abstract}
The experimental test problem of the left-right polarization-dependent lifetime asymmetry is discussed. It shows that the existing experiments cannot demonstrate the
lifetime asymmetry to be right or wrong after analyzing the measurements on the neutron, the muon and the tau lifetime, as well as the $g-2$ experiment. However, It is pointed out emphatically that the SLD and the E158 experiments, the measurements of the left-right integrated cross section asymmetry in $Z$ boson production by $e^+e^-$ collisions and by electron-electron M{\o}ller scattering, can indirectly demonstrate the lifetime asymmetry. In order to directly demonstrate the lifetime asymmetry, we propose some possible experiments on the decays of polarized muons. The precise measurement of the lifetime asymmetry could have important significance for building a muon collider, also in cosmology and astrophysics. It would provide a sensitive test of the standard model in particle physics and allow for exploration of the possible $V+A$ interactions.
\\[0.3 cm]
\\[0.2 cm]
\end{abstract}

\section{Introduction}\label{1}\noindent

\begin{figure}
\includegraphics{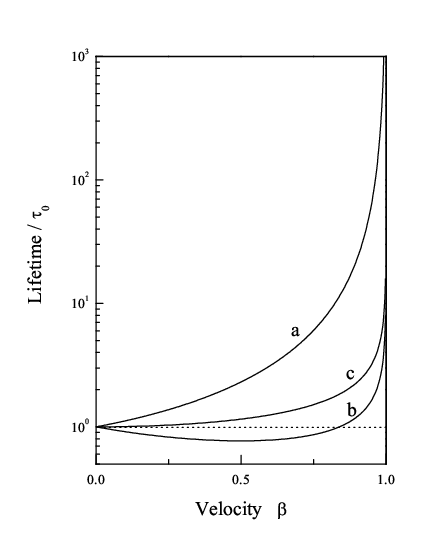}
\caption{\label{fig:lifetime-3}The lifetime as a function of the fermion velocity $\beta$. (a) The lifetime $\tau_{_{Rh}}$ of right-handed polarized fermions. (b) The
lifetime $\tau_{_{Lh}}$ of left-handed polarized fermions. (c) The lifetime $\tau$ of unpolarized fermions.}
\end{figure}

The left-right polarization-dependent lifetime asymmetry has been proposed based on theoretical analyses and concrete calculations for decays of muons within the
framework of the standard model (SM) in particle physics $^{[1-4]}$. It is shown that the lifetime of the right-handed (RH) polarized fermions is always greater than that of the left-handed (LH) ones in flight with the same speed in any inertial frame,
\begin{equation}\label{eq:1}
    \tau_{_{Rh}}=\frac{\tau}{1-\beta}\quad \hbox{and}\quad
    \tau_{_{Lh}}=\frac{\tau}{1+\beta},
\end{equation}
where $Rh(Lh)$ refers to RH(LH) helicity state, $\tau$ is the average lifetime, $\tau=\gamma\tau_{_0}=\tau_{_0}/\sqrt{1-\beta^2}$, $\beta$ is the velocity of the fermions and $\tau_{_0}$ is the lifetime of fermions in the rest frame. As shown in Fig.~\ref{fig:lifetime-3} when $\beta=0$, we find $\tau_{_{Rh}}=\tau_{_{Lh}}=\tau_{_0}$; when $\beta\rightarrow 1$, $\tau_{_{Rh}}\rightarrow\infty$, $\tau_{_{Lh}}\rightarrow\infty$. In particular, when $\beta=\frac{1}{2}$, the lifetime of the LH polarized fermions has a minimum value $\tau_{_{Lh}}=\tau_{min}=0.77\tau_{_0}$ and when $\beta\simeq 0.84$, it equals $\tau_{_0}$ again.

The lifetime asymmetry (LA) is defined as
\begin{equation}\label{eq:2}
    A\equiv\frac{\tau_{_{Rh}}-\tau_{_{Lh}}}{\tau_{_{Rh}}+\tau_{_{Lh}}}=\beta.
\end{equation}
Obviously, it is proportional to the fermion velocity $\beta$.

Similarly, for antifermions in flight, their lifetimes are
\begin{equation}\label{eq:3}
    \overline\tau_{_{Rh}}=\frac{\tau}{1+\beta},\quad\hbox{and}\quad \overline\tau_{_{Lh}}=\frac{\tau}{1-\beta}.
\end{equation}

As is well known, parity nonconservation implies that fermions exhibit asymmetrical behavior with respect to the right and the left in weak interactions. The lifetime asymmetry is a natural result of parity violation, but it is so far merely a theoretical prediction or hypothesis unverified by experimental evidence. In this paper, the reason why many experiments, including the measurement on the lifetime of the neutrons, the $\mu$ leptons, the $\tau$ leptons and the g-2 experiment, cannot test the LA will be analyzed in detail. We will introduce two experiments, the SLD and the E158 experiments, which can indirectly test the LA. Then the direct test methods of the LA will be investigated. Finally, some consequences, both theoretical and experimental, of measuring the LA will be mentioned.

\section{The measurement of the neutron and the $\mu$ lifetime}\label{2}\noindent

As seen from formula (\ref{eq:2}), the lifetime asymmetry $A=\beta$, so that it increases with the increase of velocity $\beta$ or energy of fermion, and is negligible at low energy. In order to test the LA, therefore, the fermions must be polarized and the experiments must be performed in their motion, i.e., the fermions must be moving with respect to the detector. Published neutron lifetimes were measured by using thermal or cold neutron beams $^{[5,6]}$. The energy of these beams is below 1 eV which is too low for the LA effect to be observed.

The velocity of high energy muons can exceed $\beta=0.9$, so the polarized muons are ideal objects to test the LA. Unfortunately, all experiments for measuring the
lifetime of muons were performed in the muon rest frame. In these experiments, the muons were stopped in a nuclear emulsion, sulphur, aluminum, calcium or polyethylene
target etc. In CHARM experiment $^{[7,8]}$, for example, the positive muons were produced by scattering of antineutrinos on nucleons and found to have positive helicity.
The decay process ($\mu^+\rightarrow e^+ +\nu_e+\widetilde{\nu}_{\mu}$) of muons was investigated by a measurement of the time distribution of the positrons and from
which a muon lifetime has been derived. However, It is noteworthy that the polarized muons were stopped in target material Marble (CaCO$_3$) and ``the decay positron had
to be recorded within the time window $0.94 < t < 4.89\; \mu$sec after the muon came to rest." $^{[7]}$ In the other measurement of the positive muon lifetime $^{[9]}$,
which is quoted in Review of Particle Physics $^{[10]}$, muons were produced by decaying of pions. A positive pion beam was stopped in the liquid hydrogen or sulphur
target which is known to depolarize the muons rapidly. Therefore, the stopped pions provided an unpolarized and rest muon source and the measurement has nothing to do
with the LA. TWIST Collaboration has also precisely measured the decay of polarized muons, in which the polarized muons were stopped in a $125-\mu m$ thick Mylar stopping target $^{[11]}$.

\section{The muon $g-2$ experiment}\label{3}\noindent

By the $g-2$ experiment $^{[12-15]}$, the anomalous magnetic moment of the muons was measured and the muon lifetime in flight in a circular orbit provided an
important check on the time dilation effect predicted by the special theory of relativity. Could it also provide an excellent check on the LA?

The muon beam is formed from decays ($\pi^+\rightarrow\mu^++\nu_\mu$) of a secondary pion beam. Since the pion has spin zero, the neutrino and muon must have
antiparallel spin vectors. The pions decay in flight and the decay muons project in the forward direction, thus the muons and neutrinos have parallel momentum
vectors in the laboratory frame. The neutrino has negative helicity, so that the muon must have positive helicity, i.e. it is RH polarized, and might be
partially one. According to the LA, Eq.~(\ref{eq:3}), the decay rate of RH polarized positive muons, which are antiparticles, is given by
\begin{equation}\label{eq:4}
    \lambda=\frac{1+P\beta}{\gamma\tau_0},
\end{equation}
where $P$ is the degree of muon polarization and $0\leq P\leq 1$.

However, a muon in the muon storage ring will describe a circular path and meanwhile, owing to its magnetic moment, the muon spin vector will precess about the field
direction. Because the momentum and the spin vector do not keep ``in step", the angle, $\omega_a t$, between them is a function of time and the helicity of muon does not
remain constant. Here $\omega_a$ is the angular frequency difference between the spin precession frequency and the cyclotron one. Therefore, the decay rate should
be substituted by
\begin{equation}\label{eq:5}
    \lambda=\frac{1+P\beta\cos\omega_a t}{\gamma\tau_0},
\end{equation}
it is a function of time too.

The decay formula is given by
\begin{equation}\label{eq:6}
    \frac{dN}{N}=-\lambda dt=-\frac{1+P\beta\cos\omega_a t}{\gamma\tau_0}dt.
\end{equation}
After integrating we obtain
\begin{equation}\label{eq:7}
    N=N_0e^{-\frac{P\beta}{\gamma\tau_0\omega_a}\sin\omega_a t}e^{-\frac{t}{\gamma\tau_0}}.
\end{equation}
In the g-2 experiments, $P=0.97$, $\beta =0.9994$, $\gamma\tau_0=64.4\mu s$, $\omega_a=1.4378/\mu s$ (see ref. \cite{12}), therefore we have
\begin{equation}\label{eq:8}
    \frac{P\beta}{\gamma\tau_0\omega_a}=0.01047.
\end{equation}

The exponential factor
\begin{equation}\label{eq:9}
    N_1=e^{-\frac{P\beta}{\gamma\tau_0\omega_a}\sin\omega_a t}
\end{equation}
is related to the LA. It changes with time in the range $0.98959-1.01053$ and its average value is $1.00006$. The muon lifetime is a fitted parameter in the $g-2$
experiment. One can obtain the muon lifetime and its error only if we carry out the fit to the data sets by a function in which $N_1$ is included. However, $N_1$ is not
sensitive to the fitted lifetime because it equals 1 approximately and its change with time is expected to be extremely small. Therefore, the LA cannot yet be found out
by the $g-2$ experiment under the condition described above.

In Eqs.~(\ref{eq:8}) and (\ref{eq:9}), we find $\beta/\gamma=\beta\sqrt{1-\beta^2}=0$ when $\beta=0$ and $\beta=1$. However, when $\beta=\frac{1}{\sqrt{2}}\simeq 0.707$,
it has a maximum value $\beta/\gamma=0.5$. Under this condition it might be possible to use the $g-2$ experiment as a test of the LA.

\section{The measurement of the $\tau$ lifetime}\label{4}\noindent

The lifetime of $\tau$-leptons in flight has been measured by using samples from process $e^+e^-\rightarrow Z^0\rightarrow \tau^+\tau^-$ $^{[16-18]}$. In these
measurements, three different techniques, decay length, impact parameter and impact parameter difference, were used. Directly or indirectly, all three methods involved
the measurement of an average decay length. As viewed from the measurement of the LA, however, many problems need to be discussed.

First, the polarization of $\tau$ lepton should be concretely analyzed. The polarization of $\tau$ depends on one of initial particle $Z^0$. Since $Z^0$ boson has spin 1, its helicity, $h$, can have $\pm 1$ and $0$. Along or near the incidence direction of electron beam in the collider, $\tau^-$ and $\tau^+$ must be both RH polarized when $h=+1$, LH polarized when $h=-1$, and one is RH polarized while another is LH polarized when $h=0$, according to the conservation law of angular momentum. In first and second situations, the effects of the LA of positive and negative $\tau$ leptons are counteracted one another. Only in third situation, might the effect of the LA be observed. In which if $\tau^-$ is LH polarized and $\tau^+$ is RH polarized the lifetime of $\tau$ leptons in the laboratory frame will be reduced by a factor of $(1+\beta)$ in the denominator; by contrast if $\tau^-$ is RH polarized and $\tau^+$ is LH polarized the lifetime will be lengthened by a factor of $(1-\beta)$. [see Eqs.(\ref{eq:1}) and (\ref{eq:3})]

Second, the effect of the LA might be quite small. In some experiments, the polarization of $\tau$ leptons has been measured and the results showed that $\tau^-$ was partially LH polarized and $\tau^+$ was partially RH one. However, their polarization data were quite scattered and small as well. For
example, the average polarizations were found to be only $P=0.152$ $^{[19]}$, 0.01 $^{[20]}$, 0.24 $^{[21]}$ and 0.132 $^{[22]}$, respectively. Furthermore the average
decay length or the $\tau$ lifetime will be reduced by a factor of $(1+P\beta)$ in the denominator. Therefore, the polarization must be measured precisely, and because it might be quite small the samples should be identified and treated very carefully in the fit. Unfortunately, we are unable to find more detailed information about polarization measurements in the papers of measuring the $\tau$ lifetime.

Third, there were some uncertainties in the fitting procedure. In order to obtain the average decay length, a maximum likelihood fit was applied to the observed decay
length distribution. In these experiments, however, the authenticity of the fitted results is suspicious because the measurement errors are quite big and the samples are
selected again and again. For example, in the L3 experiment only observed decay length values with an error less than 5 mm were accepted for the fit, but the average
decay lengths of the fit were 2.245 mm and 2.265 mm, respectively $^{[16]}$. In the MAC experiment, the fit included only 532 events having assigned the observed decay
length error less than 3 mm, but the average decay length was 0.777 mm $^{[17]}$. Obviously, these error values are much greater than the average values of the samples themselves. On the other hand, after each iteration, some samples were removed again $^{[18]}$. In the MAC experiment, for example, about 10 \% of the
samples were removed $^{[17]}$. It is well known that an input value is sensitive to the fitted result and plays an important role in selecting a sample being
whether or not accepted. If considering the LA in these experiments, the assigned input decay length might not be reasonable and the removed samples might be
important.

Lastly, it is more important that the tau decay is not a pure weak process. The tau decay is a multiplicity process, not like muon decay, and includes the purely leptonic decays and the semi-leptonic decays. The branching ratio of the former is about 35 \% $^{[10,16]}$ and the latter are sensitive to the strong coupling constant $^{[16]}$. Each decay mode makes different contribution to the lifetime whereas the LA is only caused by weak interactions.

To conclude, these experiments of measuring the $\tau$ lifetime cannot prove whether the LA is correct or not.

\section{Indirect demonstration of the lifetime asymmetry}\label{5}\noindent

The SLD Collaboration has measured the left-right cross section asymmetry in Z boson production by $e^+ e^-$ collisions $^{[23-27]}$
\begin{equation}\label{eq:10}
    e^-_{L,R}+e^+\rightarrow Z^0,
\end{equation}
where $L(R)$ refers to the left(right)-handed electron beam polarization. In the SLD experiments, only electrons were polarized, while positrons were unpolarized. Parity violation gives rise to asymmetrical polarization-dependent cross sections and final state angular distributions. The polarized electron beam at the SLD allows for asymmetry measurements with two different techniques. One is a left-right cross-section asymmetry and it is sensitive to the initial state coupling ($e^+e^-\rightarrow Z^0$). The other involves a forward-backward asymmetry, in which the cross sections are obtained by integrating forward and backward hemispheres separately, along with LH and RH polarized electrons. This is sensitive to the final state coupling ($Z^0\rightarrow e^+e^-,\;\mu^+\mu^-,\;\tau^+\tau^-$).

We only discuss the first technique here. The left-right cross section asymmetry is defined as
\begin{equation}\label{eq:11}
    A_{LR}=\frac{\sigma_L-\sigma_R}{\sigma_L+\sigma_R},
\end{equation}
where $\sigma_L$ and $\sigma_R$ are the $e^+ e^-$ production cross sections for $Z^0$ bosons with LH and RH polarized electron beams, respectively. They are obtained by
integrating over all kinematically allowed final states, i.e., the entire $4\pi$ solid angle $^{[25]}$. The electron beam polarization was measured by the Cherenkov detector (CKV), Polarized Gamma Counter (PGC) and Quartz Fiber Calorimeter (QFC), and the result was about 75\%. The positron beam polarization was directly measured by a M$\o$ller polarimeter in End Station A and the result, $-0.02\pm 0.07\%$, was consistent with zero. The SLD has collected about 530 K $Z^0$ events from 1992 to 1998 and got $A_{LR}=0.15056\pm 0.00239$ $^{[28]}$.

It is obvious from the experiments that $\sigma_L > \sigma_R$, i.e., the integrated cross section of LH polarized electron beams is greater than that of RH ones.
This result possesses important significance. Many phenomena of parity violation, for example, the asymmetry of the angular distribution of the particles
emitted from decay processes and the asymmetry in polarized electron-nucleon scattering, have been observed experimentally. These phenomena are related to differential
cross sections, while the $A_{LR}$ asymmetry is related to integrated one. It shows that the integrated cross section of electron-positron weak interactions is not a scalar under the space inversion. This is a veritably experimental truth though it is in conflict with our common sense. It is due to the weak coupling strength of LH chirality state to the $Z^0$ boson being stronger than that of RH chirality state in neutral weak currents. This characteristic is much more evident in charged weak currents, in which the weak coupling strength of RH chirality state to the $W^\pm$ boson is equal to zero, so that there exist only LH chirality state. Therefore, the integrated cross section induced by charged weak currents should also be not a scalar under the space inversion. The fermion's decay processes are caused by charged weak currents and the lifetime is the reciprocal of integrated cross section (i.e., decay rate), so the lifetime must not be a scalar too. It is a logical generalization of the polarization asymmetry in neutral weak currents. Consequently, in our opinion, the $A_{LR}$ measurement has already indirectly proved that the prediction about the LA should be correct.

It is noteworthy that the SLAC E158 Collaboration has also reported a precise measurement of the parity-violating asymmetry in polarized electron-electron M{\o}ller
scattering $^{[29,30]}$. The left-right asymmetry was found in neutral weak currents processes mediated by $Z^0$ exchange and result shows that the integrated cross section of LH helicity electron is greater than that of RH helicity electron. Therefore, this experiment indirectly proves the LA again.

\section{The measurement of the lifetime asymmetry}\label{6}\noindent

To test the LA, a relatively simple possibility is to measure the decays of polarized muons. In the SM, muon decay is described by a $V-A$ interaction and its decay rate is given by
\begin{equation}\label{eq:12}
    \Gamma=\frac{4\;G^2}{(2\pi)^5}\frac{1}{E_p}\int\!\frac{d^3 q}{E_q}\frac{d^3
    k}{E_{k}}\frac{d^3k'}{E_{k'}}\delta^4(p-q-k-k'){\cal F}.
\end{equation}
where $p$, $q$, $k$ and $k'$ are 4-momenta for $\mu$, $e$, $\nu_\mu$ and $\bar{\nu}_e$, respectively. Because the polarization of muons in flight must be described by
the helicity state \cite{2,3}, the decay amplitudes of LH and RH polarized muons are
\begin{equation}\label{eq:13}
    \begin{array}{l}
    {\cal F}_{Lh}=(1+\beta)[(p\cdot k')(q\cdot k)+m_\mu(e\cdot k')(q\cdot k)],\cr\noalign{\vskip2mm}
    {\cal F}_{Rh}=(1-\beta)[(p\cdot k')(q\cdot k)-m_\mu(e\cdot k')(q\cdot k)],
    \end{array}
\end{equation}
respectively, where $m_\mu$ is muon mass. If the muons are partially polarized, the above formulas should be replaced by
\begin{equation}\label{eq:14}
    \begin{array}{l}
    {\cal F}_{Lh}=(1+P\beta)[(p\cdot k')(q\cdot k)+m_\mu(e\cdot k')(q\cdot k)],\cr\noalign{\vskip2mm}
    {\cal F}_{Rh}=(1-P\beta)[(p\cdot k')(q\cdot k)-m_\mu(e\cdot k')(q\cdot k)],
    \end{array}
\end{equation}
where $P$ is the degree of muon polarization.

Neglecting electron mass and taking the simplest case of $\bm p: \bm p=p_z$, after integrating Eq.~(\ref{eq:12}), the differential decay rates of polarized muons or the
angular distributions of electrons emitted by the muons $\mu^-$ during decay are given by
\begin{equation}\label{eq:15}
  \begin{array}{l}
  \sigma_L^{\mu^-}(\theta)=\frac{\displaystyle d\Gamma_{Lh}(\mu^-)}{\displaystyle d\cos\theta}
  =\frac{\displaystyle\Gamma_0(1+P\beta)}{\displaystyle 2\gamma^3(1-\beta\cos\theta)^2}
  \Big(1+\frac{\displaystyle1}{\displaystyle3}P\frac{\displaystyle\cos\theta-\beta}{\displaystyle1-\beta\cos\theta}\Big),
  \cr\noalign{\vskip2mm}
  \sigma_R^{\mu^-}(\theta)=\frac{\displaystyle d\Gamma_{Rh}(\mu^-)}{\displaystyle d\cos\theta}
  =\frac{\displaystyle\Gamma_0(1-P\beta)}{\displaystyle 2\gamma^3(1-\beta\cos\theta)^2}
    \Big(1-\frac{\displaystyle1}{\displaystyle3}P\frac{\displaystyle\cos\theta-\beta}{\displaystyle 1-\beta\cos\theta}\Big),
  \end{array}
\end{equation}
where $\theta$ is the angle between the electron and the muon momentum, the relativistic factor $\gamma=1/\sqrt{1-\beta^2}$ and $\Gamma_0$ is the well-known
decay rate in the muon rest frame:
\begin{equation}\label{eq:16}
    \Gamma_0=\frac{G^2m^5_\mu}{192\;\pi^3}.
\end{equation}

Similarly, for polarized muons $\mu^+$, the angular distributions of positrons are given by
\begin{equation}\label{eq:17}
    \begin{array}{l}
    \sigma_L^{\mu^+}(\theta)=\frac{\displaystyle d\Gamma_{Lh}(\mu^+)}{\displaystyle d\cos\theta}
    =\frac{\displaystyle\Gamma_0(1-P\beta)}{\displaystyle 2\gamma^3(1-\beta\cos\theta)^2}
    \Big(1+\frac{\displaystyle1}{\displaystyle3}P\frac{\displaystyle\cos\theta-\beta}{\displaystyle1-\beta\cos\theta}\Big),
    \cr\noalign{\vskip2mm}
    \sigma_R^{\mu^+}(\theta)=\frac{\displaystyle d\Gamma_{Rh}(\mu^+)}{\displaystyle d\cos\theta}
    =\frac{\displaystyle\Gamma_0(1+P\beta)}{\displaystyle 2\gamma^3(1-\beta\cos\theta)^2}
    \Big(1-\frac{\displaystyle1}{\displaystyle3}P\frac{\displaystyle\cos\theta-\beta}{\displaystyle1-\beta\cos\theta}\Big),
    \end{array}
\end{equation}

It can be seen from Eqs.~(\ref{eq:15}) and (\ref{eq:17}) that the angular distributions of electrons and of positrons are of the similar form and they always tend to be
emitted along the direction which is antiparallel to the muon spin-vector, just like that in nuclear $\beta$ decay. The decay rate asymmetry can be expressed as
\begin{equation}\label{eq:18}
    A_L^{\mu^\pm}(\theta)=\frac{\sigma_L^{\mu^-}(\theta)-\sigma_L^{\mu^+}(\theta)}{\sigma_L^{\mu^-}(\theta)+\sigma_L^{\mu^+}(\theta)}
    =P\beta,\quad \hbox{and}\quad
    A_R^{\mu^\pm}(\theta)=\frac{\sigma_R^{\mu^+}(\theta)-\sigma_R^{\mu^-}(\theta)}{\sigma_R^{\mu^+}(\theta)+\sigma_R^{\mu^-}(\theta)}
    =P\beta.
\end{equation}

When $\cos\theta=\beta$, instead of Eqs.~(\ref{eq:15}) and (\ref{eq:17}) we obtain
\begin{equation}\label{eq:19}
    \begin{array}{l}
    \sigma_L^{\mu^-}(\cos\theta=\beta)=\frac{1}{2}(1+P\beta)\gamma \Gamma_0,\cr\noalign{\vskip2mm}
    \sigma_R^{\mu^-}(\cos\theta=\beta)=\frac{1}{2}(1-P\beta)\gamma \Gamma_0,\cr\noalign{\vskip2mm}
    \sigma_L^{\mu^+}(\cos\theta=\beta)=\frac{1}{2}(1-P\beta)\gamma \Gamma_0,\cr\noalign{\vskip2mm}
    \sigma_R^{\mu^+}(\cos\theta=\beta)=\frac{1}{2}(1+P\beta)\gamma \Gamma_0.
    \end{array}
\end{equation}
Obviously, this condition, $\cos\theta=\beta$, reduces the formulas of the decay rate and the decay rate asymmetry can be expressed as
\begin{equation}\label{eq:20}
    \begin{array}{l}
    A_{LR}^{\mu^-}(\cos\theta=\beta)=\frac{\displaystyle\sigma_L^{\mu^-}(\cos\theta=\beta)-\sigma_R^{\mu^-}(\cos\theta=\beta)}
    {\displaystyle\sigma_L^{\mu^-}(\cos\theta=\beta)+\sigma_R^{\mu^-}(\cos\theta=\beta)}=P\beta,
    \cr\noalign{\vskip2mm}
    A_{RL}^{\mu^+}(\cos\theta=\beta)=\frac{\displaystyle\sigma_R^{\mu^+}(\cos\theta=\beta)-\sigma_L^{\mu^+}(\cos\theta=\beta)}
    {\displaystyle\sigma_R^{\mu^+}(\cos\theta=\beta)+\sigma_L^{\mu^+}(\cos\theta=\beta)}=P\beta.
    \end{array}
\end{equation}

Because the positive muons produced by decaying of positive pions have positive helicity and the negative muons produced by negative pions have negative helicity, the asymmetry can be expressed as
\begin{equation}\label{eq:21}
    A_{LR}^{\mu^\pm}(\theta)=\frac{\sigma_L^{\mu^-}(\theta)-\sigma_R^{\mu^+}(\theta)}{\sigma_L^{\mu^-}(\theta)+\sigma_R^{\mu^+}(\theta)}
    =\frac{\displaystyle1}{\displaystyle3}P\frac{\displaystyle\cos\theta-\beta}{\displaystyle1-\beta\cos\theta}.
\end{equation}
When $\cos\theta=0$, $A_{LR}^{\mu^\pm}=-\frac{\displaystyle1}{\displaystyle3}P\beta$, when $\cos\theta=\pm 1$, $A_{LR}^{\mu^\pm}=\pm\frac{\displaystyle1}{\displaystyle3}P$.

All quantities $A_L^{\mu^\pm}(\theta)$, $A_R^{\mu^\pm}(\theta)$, $A_{LR}^{\mu^-}(\cos\theta=\beta)$, $A_{RL}^{\mu^+}(\cos\theta=\beta)$ and $A_{LR}^{\mu^\pm}(\theta)$ are proportional to the muon polarization $P$. If these quantities are not equal to zero, one will then have a definite proof of the LA.

\section{Summary and discussion}\label{7}\noindent

All existing experiments of measuring lifetime cannot be employed for testing the LA and we still need directly experimental evidence. Because pure leptonic processes obey the pure $V-A$ theory and are the simplest weak interaction ones, the muon decay, as a typical sample, is often used for researching the theory of weak interactions and also the best selection for testing the LA. In this paper it is proposed that a direct demonstration of the LA could come true by measuring the decay rate asymmetry, $A_L^{\mu^\pm}(\theta)$, $A_R^{\mu^\pm}(\theta)$, $A_{LR}^{\mu^-}(\cos\theta=\beta)$, $A_{RL}^{\mu^+}(\cos\theta=\beta)$ and $A_{LR}^{\mu^\pm}(\theta)$, of the polarized muon which should be highly polarized and moving with velocities approaching $c$ and along a straight line.

As we have discussed in section 3, of course, it might be possible to use the $g-2$ experiment as a test of the LA under condition $\beta=0.707$.

The experimental test of the LA would be more interesting and more important:

(1) The left-right cross section asymmetry $A_{LR}$ and the lifetime asymmetry $A$ are entirely similar. The former is the polarization asymmetry in neutral weak currents, theoretically and experimentally, it has been investigated widely. The latter is the polarization asymmetry in charged weak currents and its theoretical analyses have been performed by us, but the experiment on it has not yet been found in literature. The measuring the LA could observe a new parity violation phenomenon that has been overlooked for more than 50 years since the discovery of parity nonconservation in 1956-1957. It is an indispensable complement to the theory about parity nonconservation.

(2) The SLD and the E158 experiments have shown the polarization asymmetry, but they can not get a quantitative relation between the asymmetry and the polarization. When the fermions are partially polarized, the formula (\ref{eq:2}) can be rewritten as
\begin{equation}\label{eq:22}
    A\equiv\frac{\tau_{_{Rh}}-\tau_{_{Lh}}}{\tau_{_{Rh}}+\tau_{_{Lh}}}=P\beta.
\end{equation}
Obviously, the LA is equal to the product of the polarization and the velocity of decaying fermions. As indicated from Eqs.~(\ref{eq:18}), (\ref{eq:20}), (\ref{eq:21}) and (\ref{eq:22}), all the asymmetries are related to the polarization and are of the order of $\beta$, not $\beta^2$, so the LA is not a relativistic effect in the strict sense, but it is a parity violated one. The measurement of the LA might be crucial for further understanding the essence of the elusive weak interactions.

(3) ``The polarization asymmetry will play a central role in precise tests of the SM.'' $^{[31]}$ The LA given by Eq.~(\ref{eq:1}) and (\ref{eq:3}) implies a pure $V-A$ type interactions. If there is some interaction of $V+A$ type mixed in muon decay, then, in general case, the Hamiltonian can be written as
\begin{eqnarray}\label{eq:23}
    H&\sim &C_1 (V-A)(V-A)+ C_2 (V+A)(V+A)\nonumber\\
     &\sim &C_1\sum_{s,s',r,r'=1}^2\left[\;\overline{u}_{s'}(q)\;\gamma_\lambda\;(1+\gamma_5)\;v_{r'}(k')\;\right]
     \left[\;\overline{u}_r(k)\;\gamma_\lambda\;(1+\gamma_5)\;u_s(p)\;\right] \nonumber\\
     &&+\;C_2\sum_{s,s',r,r'=1}^2\left[\;\overline{u}_{s'}(q)\;\gamma_\lambda\;(1-\gamma_5)\;v_{r'}(k')\;\right]
     \left[\;\overline{u}_r(k)\;\gamma_\lambda\;(1-\gamma_5)\;u_s(p)\;\right],
\end{eqnarray}
where $s$, $s'$, $r$ and $r'$ are spin indices for $\mu$, $e$, $\nu_\mu$ and $\bar{\nu}_e$, respectively, while $C_1$ and $C_2$ are two real coefficients with constraint,
say, $C_1+C_2=1$. Because decay amplitude ${\cal F}\sim H^2$ and $(1+\gamma_5)(1-\gamma_5)=0$, after applying the analogous method in reference \cite{2}, we easily find
that the decay amplitude of LH and RH polarized muons, Eq.~(\ref{eq:13}), are modified as
\begin{equation}\label{eq:24}
    \begin{array}{l}
    {\cal F}_{Lh}=(1+\alpha\beta)[(p\cdot k')(q\cdot k)+m_\mu(e\cdot k')(q\cdot k)],\cr\noalign{\vskip2mm}
    {\cal F}_{Rh}=(1-\alpha\beta)[(p\cdot k')(q\cdot k)-m_\mu(e\cdot k')(q\cdot k)].
    \end{array}
\end{equation}
Accordingly, Eqs.~(\ref{eq:18}) and (\ref{eq:20}) should also be modified as
\begin{equation}\label{eq:25}
    A_L^{\mu^\pm}= A_R^{\mu^\pm}=\alpha P\beta
\end{equation}
and
\begin{equation}\label{eq:26}
A_{LR}^{\mu^-}(\cos\theta=\beta)= A_{RL}^{\mu^+}(\cos\theta=\beta) =\alpha P\beta,
\end{equation}
respectively, where
\begin{equation}\label{eq:27}
    \alpha=\frac{C^2_1-C^2_2}{C^2_1+C^2_2}.
\end{equation}
If $C_1=C_2$, then $\alpha=0$, there is no the LA, i.e., the parity is conserved. If $C_2=0$, then $\alpha=1$, there is no $V+A$ interaction, i.e., the parity violation
reaches its maximum. Any deviation from $\alpha=1$ would imply the muon-decay Lagrangian including $V+A$ interaction. Hence the precise measurement of the quantity $\alpha$ would determine the extent of mixing of the minor $V+A$ interaction into the major $V-A$ interaction of weak interactions. Therefore, the measurements on the LA would not only provide important and stringent tests for the SM, but also shed some light to possible new physics beyond it.

(3) Recently, Fermilab physicists hope to build a muon collider. The colliding of muons must happen before the muons decay. It is estimated roughly that `` if a muon zings along at 1.5 TeV, the time dilation of special relativity stretches its lifetime to 30 milliseconds---up from 2 microseconds when it's still. That's time enough for 500 circuits in the final ring"$^{[32]}$. If the LA is taken into account, however, the lifetime of RH polarized muons will be stretched to 156 days, while that of LH ones will be only stretched to 15 milliseconds. Therefore, it is necessary to consider the LA in the design of a muon collider.

(4) It might have consequences for cosmology and astrophysics. For example, the LA could be served as one of mechanisms to explain the ``knee" and ``ankle" in
cosmic ray spectrum $^{[33,34]}$.

\section*{Acknowledgments}

We are grateful to Dr. R. Belusevic for bringing to our attention the experimental problem of the LA and many helpful discussions.

\end{document}